# Test-Bed Based Comparison of Single and Parallel TCP and the Impact of Parallelism on Throughput and Fairness in Heterogeneous Networks

[1]Mohamed A. Alrshah, [2]Mohamed Othman

[1]Master of Computer Science, [2]PhD of Computer Science

Communication Technology and Networks, Faculty of Computer Science, UPM, Serdang, Malaysia-43400

E-mail: [1]mohammed_aid@yahoo.com, [2]mothman@fsktm.upm.edu.my

## ABSTRACT

Parallel Transport Control Protocol (TCP) has been used to effectively utilize bandwidth for data intensive applications over high Bandwidth-Delay Product (BDP) networks. On the other hand, it has been argued that, a single-based TCP connection with proper modification such as HSTCP can emulate and capture the robustness of parallel TCP and can well replace it. In this work a Comparison between Single-Based and the proposed parallel TCP has been conducted to show the differences in their performance measurements such as throughput performance and throughput ratio, as well as the link sharing Fairness also has been observed to show the impact of using the proposed Parallel TCP on the existing Single-Based TCP connections. The experiment results show that, single-based TCP cannot overcome Parallel TCP especially in heterogeneous networks where the packet losses are common. Furthermore, the proposed parallel TCP does not affect TCP fairness which makes parallel TCP highly recommended to effectively utilize bandwidth for data intensive applications.

## 1. INTRODUCTION

Parallel TCP uses a set of parallel (modified or standard) TCP connections to transfer data in an application process. With standard TCP connections, parallel TCP has been used for effectively utilize bandwidth for data intensive applications over high bandwidth-delay product (BDP) networks. On the other hand, it has been argued that a single TCP connection with proper modification can emulate and capture the robustness of parallel TCP and thus can well replace it [1].

From the implementation of parallel TCP, it is found that, the single-connection based approach (such as HSTCP) may not be able to achieve the same effects as parallel TCP, especially in heterogeneous and highly dynamic networks. The Parallel TCP achieves better throughput and performance than the single-connection based approach.

## 2. LITRATURE REVIEW

The concept of Parallel TCP is not new and its original form is the use of a set of multiple standard TCP connections. In late 80s and early/mid 90s the use of multiple standard TCP connections was exploited to overcome the limitation on TCP window size in satellite-based environments [3, 4, and 5]. But it still needs some experiments to show its efficiency and effectiveness.

The applications that require good network performance often use parallel TCP streams and TCP modifications to improve the effectiveness of TCP. If the network bottleneck is fully utilized, this approach may boosts throughput by unfairly stealing bandwidth from competing single-based TCP streams. To improve the effectiveness of TCP is much easy compared to improve the effectiveness while maintaining fairness [2]. So, TCP fairness should be measured in real test bed experiment to show its sensitivity of parallelism.

More recently, there has been a focus on improving the performance of data intensive applications, such as GridFTP [6, 7] and PSockets [8, 9]. These solutions focus on the use of multiple standard TCP connections to improve bandwidth utilization. However, these studies did not compare or identify the differences of the performance between the use of a set of multiple standard TCP connections and a single-based TCP emulating a set of multiple standard TCP connections.

Some solutions use a set of parallel connections to outperform the single standard TCP connection in terms of congestion avoidance and maintaining fairness. The approaches described in [10, 11] adopt integrated congestion control across a set of parallel connections to make them as aggressive as a single standard TCP connection. It is shown that, the approaches are fair and effective to share bandwidth. In the references [12, 13], by using a fractional multiplier the aggregate window of the parallel connections increases less than 1 packet per RTT. And only the involved connection halves its window when a packet loss is detected. Similar to Parallel TCP, pTCP uses multiple TCP Control Blocks and it shows that pTCP outperforms single connection or single TCP based approach such as MulTCP [14].

On the other hand, some solutions have the capability of using a single TCP connection to emulate the behaviour of Parallel TCP. MulTCP [14] makes one logical connection behave like a set of multiple standard TCP connections to achieve weighted proportional fairness. The recent development on high-performance TCP has resulted in TCP variants such as Scalable TCP [15], HSTCP [16], HTCP [17], BIC [18], CUBIC [19], and FAST TCP [20]. All of these TCP variants have the effect of emulating a set of parallel TCP connections.

After the fast growth of the communication devices and the increasing of network speeds, the single based TCP protocols may not be able to fully utilize the high speed network links (bandwidth). The most important thing which should be taken into account is the fairness between the connections which is a mix of single and parallel TCP connections that share the same link.

## 3. THE OBJECTIVES

In this work, real test bed experiment has been conducted to compare Single-based TCP with parallel TCP to observe the impact of using Parallel TCP rather than Single-connection based approach on increasing the performance and bandwidth utilization. Furthermore, to show the impact of using parallel TCP on TCP fairness while it shares the same bottleneck with single-based TCP connections. Jean's Fairness Index (JFI) [21], has been used in this experiment. JFI well known as the following formula:

$$(F) = \frac{(\sum_{i=1}^{N} x_i)^2}{N(\sum_{i=1}^{N} x_i^2)}$$

Where $X_i$ is the measured throughput for flow $i$, for $N$ flows in the system.

## 4. THE METHODOLOGY

In this work a comparative study has been conducted to compare between the single-based TCP and the proposed parallel TCP algorithm in terms of throughput, throughput ratio and TCP fairness based on test-bed experiment with real CUBIC version which has been implemented in the kernel 2.6.27 of Linux OpenSUSE11.1.

### 4.1 The Algorithm

In this paper, a new parallel TCP algorithm has been proposed and implemented to overcome the problems of the existent algorithms. Parallel TCP algorithms were suggested to increase the throughput of the standard TCP, because Standard TCP could not fully utilize the high-speed links, but it still suffer from fairness problem. The next two subsections will show the control flow of the sender and receiver sides' algorithm.

#### 4.1.1 Receiver-Side Model

As shown in figure 1, the main thread will keep listening for any new connection request on the determined pair of IP address and port number. If any new connection request received and authorized, the main thread will send signal to connection thread manager to establish a new connection thread. The connection thread manager will create a new connection in separated thread and will send a signal to the monitor thread to start its job, which is monitoring the active connections and gathering the data when all the connections are being ended. Each connection thread will start receiving data from the involved sender until the end of the data, and then a termination of connection signal will be sent to the monitor thread to update its state. However, the monitor thread will keep working until it receives the termination signal of the last connection from the parallelized connections set, then it will start gathering and ordering data, consequently save this data into its final form.

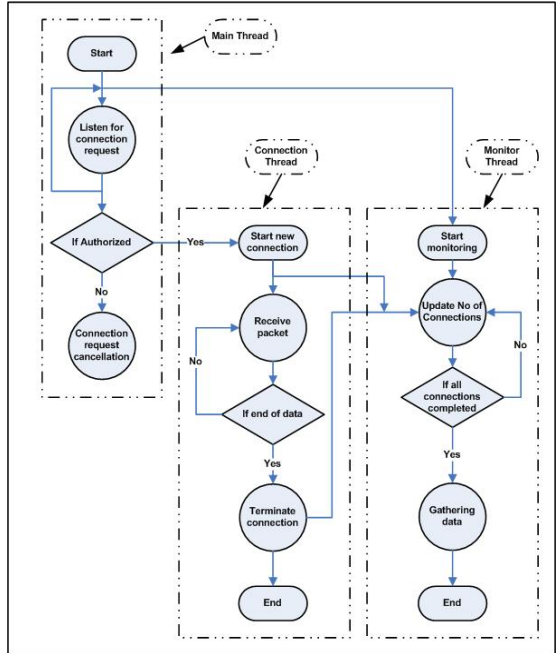

Figure 1: Receiver-Side Algorithm

### 4.1.2 Sender-Side Model

As shown in figure 2, the main thread of the sender will start working to send the data to the receiver. First, it will divide the data into chunks based on the number of connections that will be used to transfer this data. Then, it will send all the data chunks to the communication thread to start its job. Communication thread will start creating threads (connection threads) one thread for each chunk of data, and it will give different sequence numbers for each connection. Therefore, each connection thread will start data pipelining which means it will start creating packets and sending it to the receiver. When the end of the data chunk is reached, FIN packet will be sent to the receiver to start connection termination.

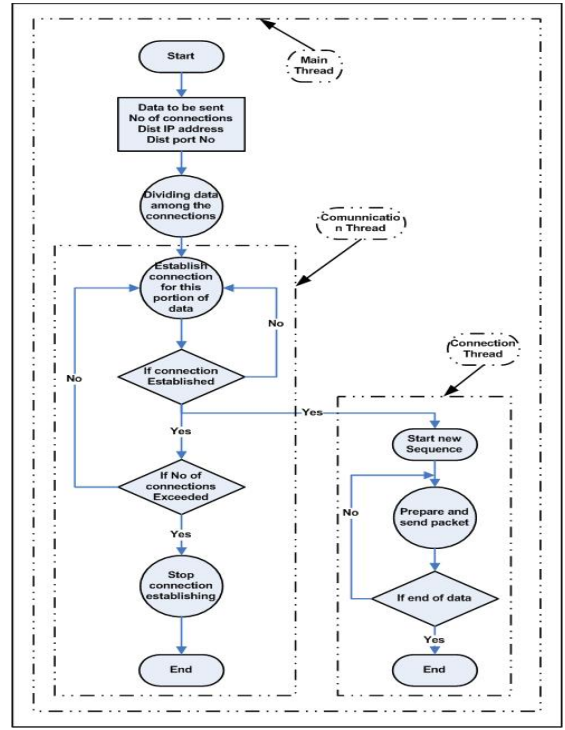

Figure 2: Sender-Side Algorithm

### 4.2 Network Topology

In this work a typical single bottleneck (dumbbell topology) with a symmetric channel and reverse path loss-free has been used as shown in figure 3. The targeted traffic has been observed in presence of the background traffic to provide the heterogeneity which is the main point of this study.

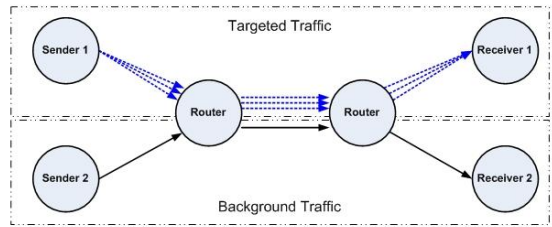

Figure 3: Network Topology

### 4.3 Performance Metrics

The Throughput, throughput ratio and TCP Fairness between the targeted and the background flows is the main point of interest in this work.

- *Throughput*: is the amount of data which have been received from the source to destination over TCP connection against the time.
- *Throughput Ratio*: is the amount of data which have been moved successfully from one source to destination in a given time period over a physical or logical link.
- *TCP Fairness*: it is known as Fairness Index which used to determine whether the applications that use the same link or connection are receiving a fair share of bandwidth or not. The Fairness Index is denoted as F, while F is a decimal value resides between 0 and 1 which represents the percentage of fairness.

## 5. HARDWARE AND SOFTWARE

In this experiment, open source C# MONO-Developer 1.0 and MONO Framework 2.0.1 has been used to build the Traffic Generator and the Traffic Collector tools on Linux OpenSUSE 11.1 operating system. This experiment has been conducted on a real network that consists of four PCs with OpenSUSE installed on it, two of them are senders and the other two PCs are receivers. One sender and one receiver have been used to provide the background traffic and the others used to provide the targeted traffic.

While in the bottleneck routers, the Linux based MikroTik RouterOS operating system version 2.9.27 has been used which provide a high level of setting to control bandwidth, routing algorithm and queue management algorithms. This routers configured to obtain the same network environment that being proposed by Qiang Fu (2007) [1].

## 6. THE RESULTS

It is well known that, in single-based TCP, AIMD algorithm reduces its congestion window to the half when packet loss detected, that leads to decrease the total throughput to the half as well. Thus, it leads to insufficient bandwidth utilization. Contrarily, in the proposed Parallel TCP algorithm, the timeout which occurs in one of parallelized flows will not affect the congestion window of the other flows in the same set consequently will quietly affect the aggregated congestion window.

For instance, suppose that we have parallel of five TCP connections, and timeout takes places on two flows, this will result the reduction of around 2/10 from the total throughput of this parallel connection. The value of 2/10 represents the half of congestion window of the affected flows.

However, the results of this work reveals that, the single-based TCP can not emulate the parallel TCP in terms of performance Throughput and throughput ratio, thus cannot replace it especially in heterogeneous networks. Figure 4 and figure 5 show that, the increasing of the number of parallel TCP connections that used for one application process increases the throughput of that application, which makes the use of the proposed parallel TCP algorithm highly recommended for data intensive applications to achieve better performance especially when high speed network links are used.

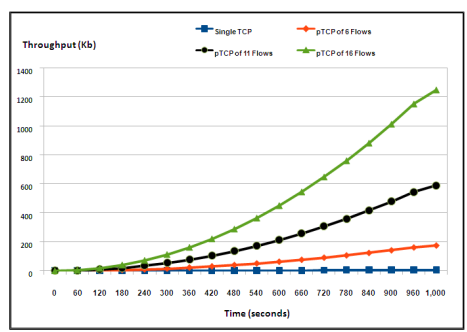

Figure 4: Performance Throughput

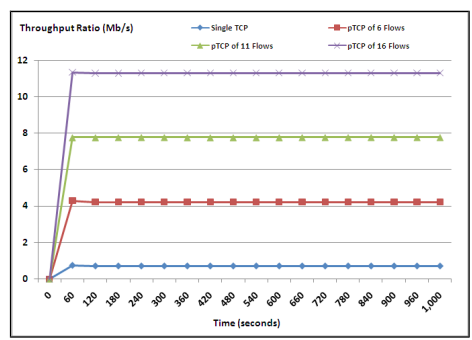

Figure 5: Throughput Ratio

Since the parallelism occurs in the application layer that divides the data into small chunks and provide them to the lower layers in order to establish an independent connections for them. These flows will behave independently in the lower layers of the network and it will result positive impact on TCP fairness as shown in figure 6 and figure 7.

## 7. CONCLUSION

In this paper, it has been demonstrated that the proposed Parallel TCP outperforms single-based TCP in high speed and highly dynamic networks and it (parallel TCP) produced better throughput than the single-based TCP while it maintains the fairness with single based flows.

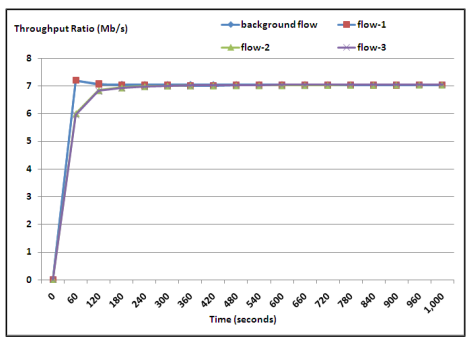

Figure 6: Throughput of the flows that share bottleneck

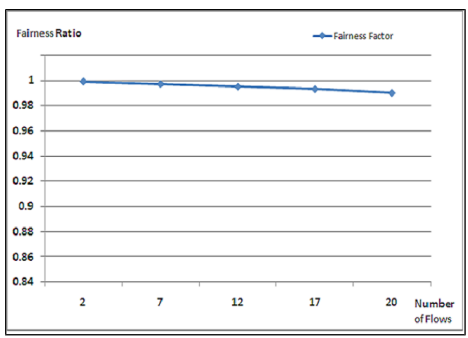

Figure 7: Fairness index of the proposed algorithm with single TCP flow

## 8. FUTURE WORK

In this work, the comparison was conducted using CUBIC and it was not good enough to say that, parallel TCP is better than single-based TCP but in the nearest future, there is an intention to conduct a new experiment which will involve some of high speed TCP variants such as HTCP, HSTCP, Scalable TCP and NewReno, and the measurements of loss ratio as well.

## 9. REFRENCES:

## BIOGRAPHY:

**Mohamed A. Alrshah** received his BSc degree in
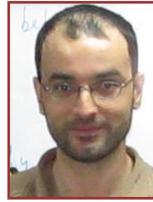
Computer Science from Naser University - Libya, in 2000, and his MSc degree in computer networks and information technology in May 2009 from University Putra Malaysia. His interests are in Parallel Computing and Computer Networks.

**Associated Professor Dr. Mohamed Othman**
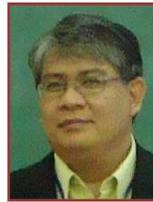
received his PhD in 1999 from University Kebangsaan Malaysia, UKM. He is working now at University Putra Malaysia. He was recently appointed as a deputy dean of faculty of computer science. He has published many articles in cited journals. His current research interests concern on Parallel and Distributed Algorithms, Grid Computing, Scientific Computing, High Speed Computer Network.